\documentclass[pdflatex,sn-nature, iicol]{sn-jnl}
\usepackage{graphicx}
\usepackage{multirow}
\usepackage{amsmath}
\usepackage{amsfonts}
\usepackage[title]{appendix}
\usepackage{booktabs}
\usepackage{algorithmicx}
\usepackage{listings}
\usepackage{siunitx}
\usepackage{mathtools}

\theoremstyle{thmstyleone}

\theoremstyle{thmstyletwo}

\theoremstyle{thmstylethree}

\begin{document}

\title[Article Title]{The Saturable Electronic Reluctance Switch: Switchable low-power and low-noise generation of magnetic fields using permanent magnets}

\author[1]{\fnm{P.D.} \sur{Taylor-Burdett}}

\author[1,2]{\fnm{C.A.} \sur{Burhan}}

\author[1]{\fnm{S.} \sur{Mason}}

\author[3]{\fnm{F.R.} \sur{Lebrun-Gallagher}}

\author[1,3]{\fnm{S.} \sur{Weidt}}

\author*[1,3]{\fnm{W.K.} \sur{Hensinger}}\email{w.k.hensinger@sussex.ac.uk}

\affil[1]{\orgname{Sussex Centre for Quantum Technologies, University of Sussex}, \orgaddress{ \city{Brighton}, \postcode{BN1 9RH}, \country{UK}}}

\affil[2]{\orgdiv{Quantum Information Science and Technologies Centre for Doctoral Training}, \orgname{University of Sussex}, \orgaddress{ \city{Brighton}, \postcode{BN1 9RH}, \country{UK}}}

\affil[3]{\orgname{Universal Quantum Ltd}, \orgaddress{\street{Gemini House, Mill Green Business Estate}, \city{Haywards Heath}, \postcode{RH16 1XQ}, \country{UK}}}

\abstract{Across many areas of science, there is a need to generate magnetic fields that are both ultra-stable and switchable “on” and “off”. Current-carrying wire configurations are switchable but are susceptible to current noise. Existing current-controlled approaches to switching the field produced by a permanent magnet involve altering the magnet’s magnetisation, which typically requires large field pulses and produces excessive power dissipation in high frequency applications. We present a hybrid technique to switch the field of any arbitrary magnet through use of a non-linear ferromagnetic circuit, named the ‘Saturable Electronic Reluctance Switch’ (SERS). The circuit achieves a linear and monotonic ramp of the magnetic field up to a current threshold, above which the field becomes constant. Crucially, the applied current has minimal influence on the magnetic field stability and demagnetisation of the magnet is avoided. The power dissipated in each switching cycle is expected to be many orders of magnitude less than for existing permanent magnet switching approaches. SERS is also robust to fabrication errors, suppressing noise in the control current by several orders of magnitude in a non-ideal device. To illustrate its application, a SERS-driven device is proposed for generating ultra-stable magnetic field gradients in a scalable trapped-ion quantum computer. We find this device offers an order of magnitude reduction in power dissipation compared to state-of-the-art current carrying wires, while reducing magnetic field noise originating from current fluctuations by up to five orders of magnitude.}

\keywords{magnetic field switching, hybrid magnetic devices, ferromagnetic circuits, low-noise magnetic fields, magnetic field gradients, trapped ions, quantum technologies, atomic and molecular physics}

\maketitle

\clearpage
\subsection*{Introduction}\label{sec1}

Across many applications that require generation of static magnetic fields, there is a recurring trade-off between achieving a field that is highly stable and simultaneously switchable. 

Important cases are found in the fields of atomic, molecular and optical physics, where fluctuations in applied magnetic fields are a significant consideration. One example in quantum information processing involves trapped-ion quantum computing using the Magnetic Gradient Induced Coupling (MAGIC) scheme \cite{Mintert2001, Nagies2025} with Zeeman-sensitive qubits in a quantum charged coupled device (QCCD) type architecture, yielding a promising avenue for scalable quantum computing \cite{Lekitsch2017}. Here, fluctuations in the applied static magnetic fields limit the achievable coherence times and gate fidelities \cite{Valahu2022}. Permanent magnets produce exceptionally low-noise fields, enabling record-breaking coherence times in trapped ions \cite{Wang2021,Ruster2016}. Yet, a large-scale QCCD architecture will likely require zeroing the applied magnetic fields during fast ion-transport through gate zones to prevent qubit state dephasing \cite{Walther2011, LeBrun2021}. Current-carrying wires (CCW) enable switchable magnetic fields \cite{siegele2022}, but suffer from technical noise and necessitate the development of sophisticated ultra-low noise current supplies \cite{merkel2019,Liu2023,Morzyski2023,Xu2019}. This apparent dichotomy between stability and bandwidth can be found in experiments involving atom traps \cite{Folman2002,Keil2016}, quantum metrology \cite{Bonus2025,Wojciechowski2010}, atomic clocks \cite{Szmuk2015,Wojciechowski2010}, Bose-Einstein condensates \cite{Fernholz2008}, quantum interferometry \cite{Elahi2025,NarasimhaMoorthy2025}, Feshbach resonances \cite{Borkowski2023}, penning traps \cite{McMahon2020,Lacy2022} and nanoMRI \cite{ScheinLubomirsky2025}, to name a few.

A few non-mechanical methods exist to effectively switch the field of a permanent magnet. Degaussing \cite{Wu2020} and electro-permanent magnet assemblies \cite{Knaian2010,Ding2025} involve cycling the magnetisation of hard and semi-hard ferromagnets via current pulses. These techniques preserve the field stability of the magnets in the ON/OFF states, but typically require large current pulses to drive the magnets into saturation and, due to their large hysteresis, are prone to excessive power dissipation in high frequency applications. An alternative method which may avoid demagnetisation of the magnet is to superpose its field with that of a CCW, temporarily zeroing the field \cite{Knaian2010}. However, matching the permanent magnet field requires a large ampere-turn product, and the resulting field is subject to current noise when the CCW is energised.

\begin{figure}[h]
\centering
\includegraphics[width=0.9\columnwidth]{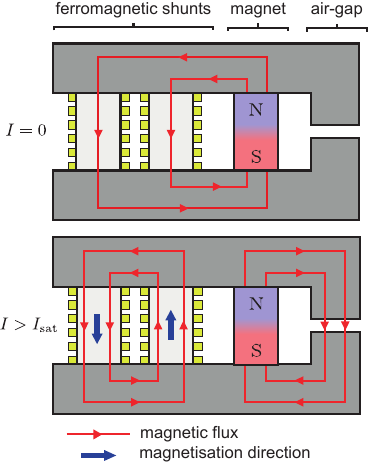}
\caption{Cross-section of a Saturable Electronic Reluctance Switch magnetic circuit, illustrating the operating principle in the low-field ``OFF'' (top) and high-field ``ON'' (bottom) states. Switching is actioned by driving two shunt cores into full saturation using anti-parallel solenoids. In the OFF-state, flux from a permanent magnet is directed through shunts with high magnetic permeability. When a driving current around the shunts is switched on, the shunts lose their permeability and the flux is instead directed towards an air-gap output region. By magnetising the shunts along a single axis, full saturation is achieved, ensuring permeability remains constant in the ON-state. The anti-parallel solenoid orientation ensures zero net contribution of the solenoid fields at the air-gap, eliminating current dependency.}
\label{fig:sers-diagram-basic}
\end{figure}

\begin{figure*}[h]
\centering
\includegraphics[width=0.9\textwidth]{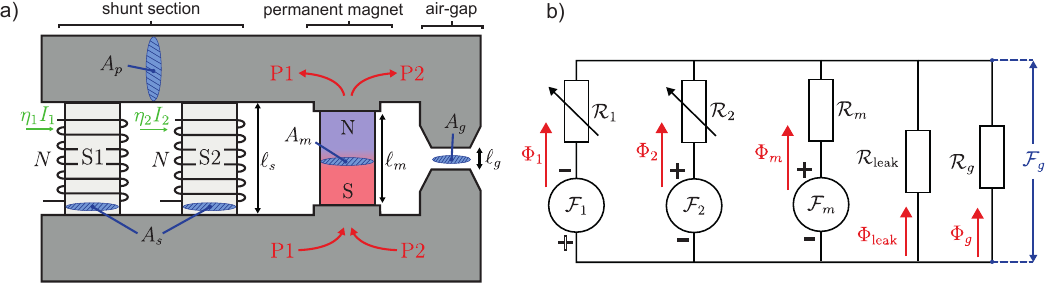}
\caption{a) Schematic of a simple Saturable Electronic Reluctance Switch (SERS) device. The device contains a primary soft ferromagnetic core of cross-sectional area $A_\text{p}$ that provides two paths, P1 and P2, for the magnetic flux, $\Phi_{m}$, of a permanent magnet of length $\ell_{m}$ and area $A_{m}$. P1 contains two high-permeability shunts, S1 and S2, of area $A_{s}$ and length $\ell_{s}$. Each shunt is tightly wound with an $N$-turn solenoid carrying a current $\eta_{n} I$, where $\eta_{n}=(-1)^{n}$. P2 contains an air-gap of length $\ell_{g}$ and area $A_{g}$, where the experiment requiring a switchable field is located. b) An idealised magnetic circuit model of the device, excluding primary core reluctance.}
\label{fig:sers-diagram}
\end{figure*}

In this work we present a non-mechanical, low-power and low-noise method of switching the magnetic field generated by a magnet at the air-gap of a ferromagnetic core. Called the Saturable Electronic Reluctance Switch (SERS), the device exploits the non-linearity of ferromagnets, magnetisation directionality and symmetry to switch the magnetic field of a permanent magnet between high-field (ON) and low-field (OFF) states, as shown in Fig. \ref{fig:sers-diagram-basic}. This relatively simple magnetic circuit creates an output field that is remarkably insensitive to the applied current. In addition, demagnetisation of the magnet is avoided, enhancing output repeatability between switching cycles and reducing power dissipation.

\subsection*{Saturable Electronic Reluctance Switch}\label{sec2}

A key property of soft ferromagnets is their ability to guide magnetic flux. In magnetic circuit terms, they provide a low-reluctance path through space, with reluctance defined by $\mathcal{R} = \ell/\mu A$, where $\ell$ and $A$ the respective length and cross-sectional area of the path and $\mu$ the medium's permeability. At low magnetisation, soft ferromagnets can display permeabilities as high as $10^{5}$ times that of vacuum, $\mu_{0}$ \cite{Jiles1991}. Such permeabilities arise from the ease with which the ferromagnet's internal magnetic moments are able to reconfigure to an applied field. However, the density of magnetic moments within the ferromagnet provide a fundamental limit to its magnetisation, leading to a maximum saturation flux density, $B_{\textrm{sat}}$. When fully saturated along a single axis, no further changes in magnetisation along the same axis can occur. Any additional field applied along this axis is subject to a constant permeability of $\mu=\mu_{0}$. The magnetisation stability obtained in full single-axis saturation is the key principle exploited by the SERS method. 

A basic configuration for the SERS device is shown in Fig. \ref{fig:sers-diagram}(a) with its idealised magnetic circuit shown in Fig. \ref{fig:sers-diagram}(b). The circuit contains three key elements: a magnet, an air-gap section and a shunt section. These are connected in parallel by a high permeability ferromagnetic ``primary'' core. The magnet, e.g. a permanent-magnet, a superconducting magnet or a solenoid, provides the air-gap field to be switched. The air-gap contains the experimental operating region where the switchable magnetic field is required. The shunt section consists of a series of solenoid-wound high permeability ferromagnetic cores configured to magnetise in anti-parallel directions. This section magnetically connects the two poles of the primary core, effectively closing the circuit for the magnet. The primary core is designed to present a negligible reluctance to the magnetic circuit throughout operation. In its simplest form, the shunt section is composed of two identical shunts. 

Following the example from Fig. \ref{fig:sers-diagram}(a), the SERS device operates by toggling the magnetic flux from the magnet between path P1, containing the shunt section, and path P2, containing the air-gap. When no current is applied to the solenoids, we approximate P1 to have a reluctance of $\mathcal{R}_{\text{off}}=\ell_{s}/2\mu_{i} A_{s}$, where $\mu_{i}$ is the initial permeability of shunt material. The reluctance of P2 is that of the air-gap, given by $\mathcal{R}_{g}=\ell_g/\mu_{0}A_{g}$. In addition, some magnetic flux will leak through the remaining volume of space between the poles of the primary core. We represent this leakage by a lumped parallel reluctance $\mathcal{R}_{\textrm{leak}}$. Since $\mu_{i} \gg \mu_{0}$ such that $\mathcal{R}_{\textrm{off}}\ll\mathcal{R}_{g}$, the proportion of the magnet's flux, $\Phi_{m}$, directed to the air-gap, $\Phi_{g}$, is then given as a proportion of the parallel reluctances as 

\begin{equation}
\Phi_{g} = \frac{\mathcal{R}_{\textrm{off}}\mathcal{R}_{\textrm{leak}}}{\mathcal{R}_{\textrm{off}}\mathcal{R}_{g}+\mathcal{R}_{g}\mathcal{R}_{\textrm{leak}}+\mathcal{R}_{\textrm{off}}\mathcal{R}_{\textrm{leak}}}\Phi_{m}\approx0.
\label{SERSeq3}
\end{equation}
Minimal flux exists at the air-gap and the switch is in the OFF-state.

To ensure the primary core and shunts are not saturated by the magnet's flux in the OFF-state, we impose the following general design condition:

\begin{description}

\item \textbf{Condition D1.} Given a magnet of area $A_{m}$ sustaining a magnetic flux density $B_{m}$, the primary core and shunts must satisfy

\begin{equation}
A_{m}B_{m}<\sum_{n}{A_{s, n}B_{\textrm{sat}, n}}\ll A_{p}B_{\textrm{sat}, p},
\label{SERSeq2}
\end{equation}

where $A_p$ is the smallest cross sectional area of the primary core and $B_{\textrm{sat}, p}$ its saturation flux density and, for a set of shunts $\{S(1)\dots S(n)\}$, where $n\geq2$, $A_{s,n}$ and $B_{\textrm{sat}, n}$ are the areas and saturation flux density of each shunt core.

\end{description}

When a current of $I_{\textrm{on}}>I_{\text{sat}}$ is applied to the solenoids, where $I_{\text{sat}}$ is the current threshold required to drive both shunts into full saturation, both shunts reach their maximum reluctance value given by $\mathcal{R}_{\textrm{on}}=\ell_{s}/2\mu_{0} A_{s}$ and hence,

\begin{equation}
\Phi_{g} = \frac{\mathcal{R}_{\textrm{on}}\mathcal{R}_{\textrm{leak}}}{\mathcal{R}_{\textrm{on}}\mathcal{R}_{g}+\mathcal{R}_{g}\mathcal{R}_{\textrm{leak}}+\mathcal{R}_{\textrm{on}}\mathcal{R}_{\textrm{leak}}}\Phi_{m} \sim \Phi_{m}.
\label{SERSeq4}
\end{equation}

The system is now in the ON-state. To maximise delivery of magnetic flux to the air-gap, we ensure the shunt reluctance is greater than the air-gap reluctance by imposing the design condition:

\begin{description}
\item \textbf{Condition D2.} For an air-gap of effective length $\ell_{g}$ and area $A_{g}$, the shunt core lengths, $\ell_{s,n}$, and areas must satisfy

\begin{equation}
\frac{\ell_{g}}{A_{g}} <  \frac{\ell_{s,n}}{A_{s,n}}.
\label{SERSeq1}
\end{equation}
\end{description}

Crucially, since both shunt cores and solenoids are identical and magnetised in anti-parallel directions, their net contribution to $\Phi_{g}$ is zero. Generally, this is achieved by imposing the design conditions:

\begin{description}

\item \textbf{Condition D3.} To achieve cancellation of the solenoid fields in the air-gap in the ON-state, the number of turns $N_{n}$, current $I_{n}$ and area $A_{\textrm{sol},n}$ of the shunt solenoids must satisfy

\begin{equation}
    \sum_{n}{\frac{\eta_{n}N_{n}A_{\textrm{sol},n}I_{n}}{\ell_{s,n}}}=0,
\end{equation}
where $\eta_{n}=(-1)^{n}$ determines the orientation of the solenoid. In addition, each solenoid must have length $\ell_{\textrm{sol},n} = \ell_{s,n}$ and must be tightly wound.

\end{description}

\begin{description}

\item \textbf{Condition D4.} To achieve cancellation of the core fields in the air-gap in the ON-state, we impose

\begin{equation}
    \sum_{n}{\eta_{n}A_{s,n}B_{\textrm{sat},n}} = 0,
\end{equation}
where $A_{s,n}$ and $B_{sat,n}$ are the area and saturation flux density of each shunt core.
 
\end{description}

In the ON-state, magnetisation changes within the shunt are avoided by saturating the cores along the same axis that shunts the magnet's flux. The air-gap field is therefore insensitive to variations on the applied current, $\delta I(t)$, provided $I_{\textrm{on}}+\delta I(t)>I_{\text{sat}}$. When the current is removed, the permeability of the shunts is restored and the system returns to the OFF-state. Through this method, the largest demagnetising field experienced by the magnet is its self-demagnetising field in the ON-state, as determined by the air-gap geometry. Only the soft ferromagnetic materials change magnetisation in each switching cycle. Therefore, power dissipation due to hysteresis losses is expected to be many orders of magnitude lower than in electro-permanent magnet and permanent magnet degaussing methods.

\begin{figure*}[h]
\centering
\includegraphics[width=0.76\textwidth]{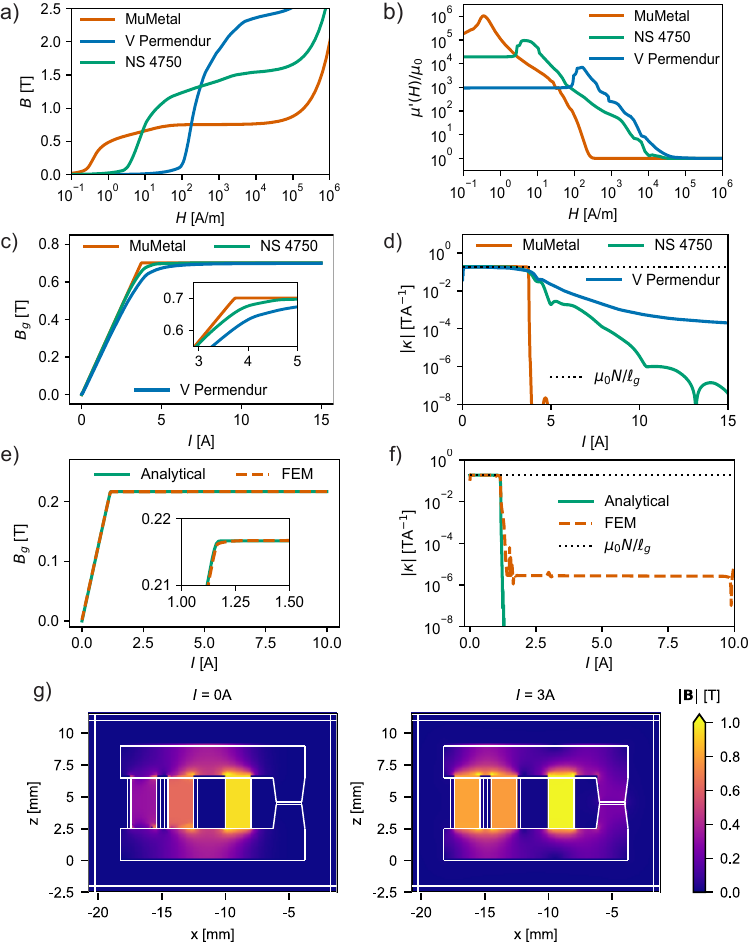}
\caption{Numerical and finite element modelling (FEM) results of a SERS circuit. Shown are the a) representative B-H curves and b) differential relative permeability of three tested shunt materials. The numerically solved air-gap B-field and calculated current sensitivity, $\kappa=dB_{g}/dI$, as a function of applied current are shown in c) and d) respectively. Switching is observed by the step-like transition from a linear ramp in the air-gap field to a nearly-constant field value as the applied current is increased. For an ideal SERS circuit, the initial linear ramp in air-gap field is given by $\kappa_{\textrm{off}}=\mu_0 N/ \ell_g$, as highlighted with a dashed line. The same two parameters are calculated for a non-linear 3D FEM simulation of the MuMetal-shunt SERS circuit and fitted with the analytical model. These are shown in e) and f). g) 3D FEM simulation of the SERS operation in the OFF (left) and ON (right) states, showing absolute magnetic flux values in each region. In the ON-state, the magnetised shunts form a visible magnetic circuit loop and a low-field region forms between the shunts and magnet, indicating isolation of the shunt region.}
\label{fig:sers-comsol}
\end{figure*}

To demonstrate the switching profile of the two-shunt SERS circuit, we derive the full circuit equation and numerically solve it to determine the air-gap flux density, $B_{g}(I)$, over a current range of \SI{0}{\ampere} to \SI{15}{\ampere} for an idealised example (see Methods). In this ideal case, all design conditions are satisfied and there is no leakage or primary core reluctance. Three different shunt materials are modelled: MuMetal\textsuperscript{\copyright}, Nickel steel (NS) 4750 and Vanadium (V) Permendur, illustrating how the material's BH-curve affects switching. Figures \ref{fig:sers-comsol}(a) and (b) illustrate the respective BH-curves and calculated differential permeabilities for each alloy. Fig. \ref{fig:sers-comsol}(c) illustrates the numerically obtained $B_{g}(I)$. Fig. \ref{fig:sers-comsol}(d) shows the air-gap current sensitivity
\begin{equation}
\kappa(I) = \frac{dB_{g}(I)}{dI}.
\end{equation}
The MuMetal shunts exemplify an ideal switch performance, where $\kappa$  falls below \SI{1E-8}{\tesla\per\ampere} and becomes numerically indistinguishable from zero. The other shunt alloys, with significantly higher saturation fields, exhibit slower and less well-defined switching. 

Defining $H_{\textrm{sat}}$ as the field strength required for full-saturation of the shunt material, we derive  
\begin{equation}
I_{\textrm{sat}} = \frac{1}{N}\left(H_{\textrm{sat}}\ell_{s}+B_{g,\textrm{sat}}\ell_{g}/\mu_{0}\right)
\end{equation}
for the switching current, where $B_{g,\textrm{sat}}$ is the final air-gap flux density in the ON-state (see Methods). An optimal shunt material can be defined as one in which $H_{\text{sat}}\ll B_{g,\text{sat}}\ell_{g}/\mu_{0}\ell_{s}$, such that
\begin{equation}
I_{\text{sat}}\approx \frac{\ell_{g}}{\mu_{0}N}B_{g,\text{sat}}.
\label{SERSeq14}
\end{equation}
This is equal to the current that would otherwise be required for a shunt solenoid to magnetise the air-gap to $B_{g,\text{sat}}$. Below saturation, the current sensitivity is equal to  $\kappa_{\textrm{off}}=\mu_{0}N/\ell_{g}$. For the example circuits, we calculate $I_{\textrm{sat,\, MuMetal}}=$\SI{3.8}{\ampere}, $I_{\textrm{sat,\, NS \,4750}}=$ \SI{9.1}{\ampere} and $I_{\textrm{sat,\, V\, Permendur}}=$ \SI{44}{\ampere}. 

We corroborate these results against a system with flux leakage and finite primary core reluctance by simulating the MuMetal case using 3D non-linear finite element method (FEM) software (see Methods). The obtained $B_{g}(I)$ and $\kappa(I)$ are shown in Fig. \ref{fig:sers-comsol}(e) and \ref{fig:sers-comsol}(f) respectively. The analytical model provides a close fit to the FEM results. However, the FEM simulation yields a finite $\kappa_{\textrm{on}}$, indicating a departure from the ideal case. Fig. \ref{fig:sers-comsol}(g) shows the magnetic field redistribution that occurs between the OFF and ON-states. In the ON-state, a low-field region appears between the shunts and the magnet, indicating magnetic isolation between these parts.

\subsection*{SERS stability and repeatability}

In the ideal SERS circuit, current sensitivity of the output field vanishes in the ON-state and field stability is limited only by the intrinsic magnetisation noise and drift of the magnet and ferromagnetic materials, e.g. thermal magnetisation fluctuations \cite{Durin1993} and the magnetic after-effect drift \cite{Preisach2017}. This condition requires full saturation of the shunts and perfect cancellation of the solenoid fields within the primary core. The former is achievable by appropriate choice of materials and optimisation of their magnetic properties, e.g. through magnetic heat-treatment \cite{Arpaia2021}. The latter is limited by fabrication tolerances, magnetic field leakage from the coils and finite primary core reluctance. For example, in the two-shunt design, failing to meet condition D3 yields a finite current sensitivity. If we define a switching ratio, $\alpha=|\kappa_{\textrm{on}}/\kappa_{\textrm{off}}|$, it can be shown that (see Methods) 
\begin{figure*}[h]
\centering
\includegraphics[width=0.9\textwidth]{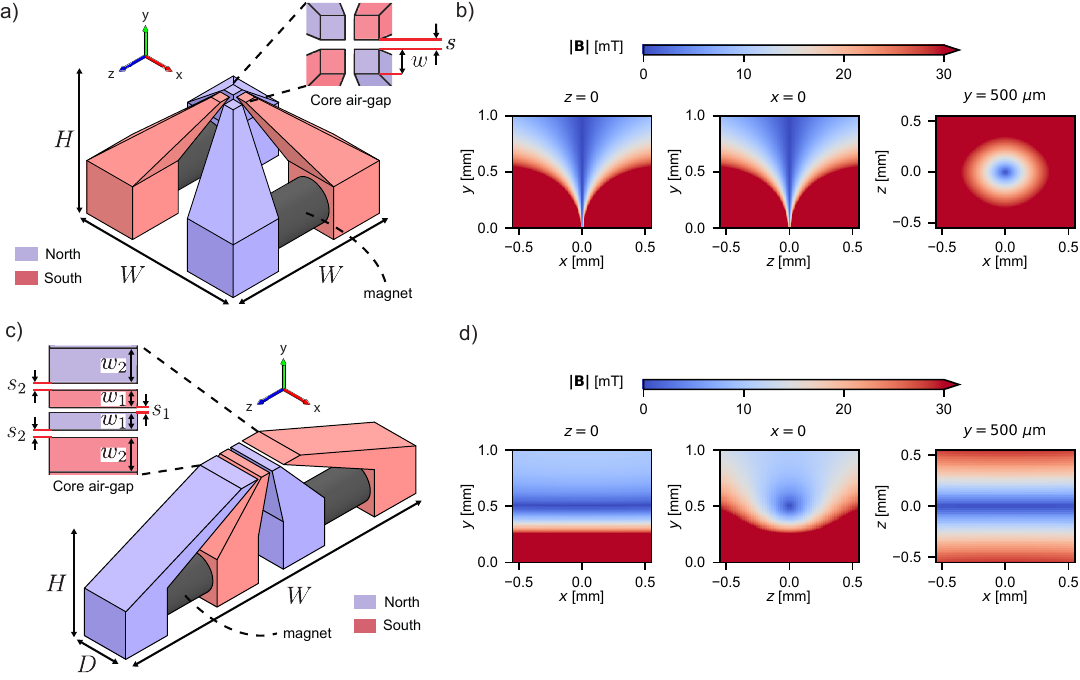}
\caption{Ferromagnetic fringe effect gradient-projecting core structures. Magnetic polarity is indicated by colour (blue/red). a) Ring core arrangement (RCA) structure, with prototype parameters of $s=$ \SI{0.2}{\milli\metre}, $w=$ \SI{0.5}{\milli\metre}, $H=$ \SI{5.5}{\milli\metre} and $W=$ \SI{8}{\milli\metre}. b) FEM simulated magnetic field magnitude in all three planes for the RCA core after magnetisation with cylindrical permanent magnets of remanent flux density $B_{r}=$\SI{1}{\tesla}, length $\ell_{m}=$\SI{3}{\milli\metre} and radius $r_{m}=$\SI{1}{\milli\metre}. c) Linear core arrangement (LCA) structure, with prototype parameters of $s_{1}=$ \SI{0.15}{\milli\metre}, $s_{2}=$ \SI{0.2}{\milli\metre}, $w_{1}=$ \SI{0.5}{\milli\metre}, $w_{2}=$ \SI{1}{\milli\metre}, $D=$ \SI{1}{\milli\metre}, $H=$ \SI{5.5}{\milli\metre} and $W=$ \SI{17.5}{\milli\metre}. Illustrated for each core is an example placement of cylindrical magnets, e.g. permanent magnets or core-filled solenoids. d) FEM simulated magnetic field magnitude in all three planes for the LCA core, magnetised by the same set of permanent magnets.}
\label{core-geometries}
\end{figure*}

\begin{equation}
    \alpha=\mu_{0}\mathcal{R}_{t}\left|\frac{A_{\textrm{sol},2}}{\ell_{2}}-\frac{A_{\textrm{sol},1}}{\ell_{1}}\right|,
    \label{methEq-finitekappa}
\end{equation}

where $A_{\textrm{sol},1}$ and $A_{\textrm{sol},2}$ are the shunt solenoid areas, $\ell_{1}$ and $\ell_{2}$ the shunt lengths and $\mathcal{R}_{t}$ is the total reluctance between the two poles of the primary core, estimated as
\begin{equation}
    \mathcal{R}_{t}\approx\left(\frac{1}{\mathcal{R}_{\textrm{on}}}+\frac{1}{\mathcal{R}_{m}}+\frac{1}{\mathcal{R_{\textrm{leak}}}}+\frac{1}{\mathcal{R}_{g}}\right)^{-1}.
\end{equation}
For the example SERS circuit including leakage, a 10\% error in the $A_{\textrm{sol},n}/\ell_{s}$ ratio of a single shunt yields an $\alpha$ of $1.1\times10^{-4}$. If the solenoid length is smaller than that of the shunt, flux leakage will occur leading to reduced cancellation in the primary core and stray fields. A FEM simulation of the example SERS device where the length of a single solenoid is reduced by 1\% and 10\% of $\ell_{s}$ yields $\alpha$ of approximately $2\times10^{-5}$ and $7\times10^{-5}$ respectively (see Methods). Another limiting factor for $\alpha$  is the finite primary core reluctance. For the example SERS device, a primary core permeability of $\mu_{r}=10^{4}$ theoretically limits $\alpha$ to approximately $8\times 10^{-6}$ (see Methods). This limit is mitigated by making the shunts equidistant from the air-gap. Where D4 is not satisfied, a constant offset, $\Delta B_{g}$, in the output field will result. In the two-shunt case,
\begin{equation}
    \Delta B_{g}=A_{s,2}B_{\textrm{sat},2}-A_{s,1}B_{\textrm{sat},1}.
\end{equation}
However, this offset does not lead to increased current dependency in the generated field.

While no significant magnetisation changes occur to the magnet during switching, the shunts and primary core undergo large excursions of their BH-loops that are subject to hysteresis. However, since the switch toggles between two stable magnetisation states, ON/OFF-state repeatability may be protected by the magnetic accommodation effect which results in convergence to a highly repeatable minor-loop on the BH-curve \cite{Tellini2009}. By cycling the switch a number of times, it may be conditioned to create a repeatable output even in the presence of hysteresis.

\subsection*{Fringe Field Quadrupole}\label{sec3}

To generate magnetic field gradients using SERS, we present the Fringe Field Quadrupole (FFQ) cores. These consist of four core poles of alternating magnetic polarities that converge into a series of air-gaps that generate a quadrupolar fringe-field. Generally, two magnets are required to magnetise the FFQ poles. The first FFQ geometry, named the Ring Core Arrangement (RCA), is shown in Fig. \ref{core-geometries}(a). The fringe-field shaping parameters are $s$ and $w$, which enable gradient magnitude optimisation at a given operating height (y-axis), $h$, above the core surface. We simulate the fields created by a prototype of the RCA core when magnetised by two permanent magnets (see Methods). The core is optimised for $h=$ \SI{500}{\micro\metre}, yielding a magnetic field gradient along the z-axis of $\partial |B|/ \partial z =$ \SI{104.7(4)}{\tesla\per\metre}. The resulting fringe-field topology is shown in Fig. \ref{core-geometries}(b), featuring a central magnetic field nil that extends indefinitely along the y-axis and a radially-symmetric magnetic field gradient in the xz-plane about the nil.

Fig. \ref{core-geometries}(c) illustrates an alternative FFQ geometry, named the Linear Core Arrangement (LCA). Fig. \ref{core-geometries}(d) illustrates the simulated magnetic fields of an LCA core prototype. A magnetic field nil exists along the x-axis with maximal gradient obtained along the z-direction.  In this design, the fringe-field shaping parameters are $s_{1,2}$ and $w_{1,2}$, which determine both the gradient magnitude and nil height above the core, and $D$, which determines the nil length along the x-axis. These parameters are chosen to yield a nil located at $h=$ \SI{500}{\micro\metre} with an obtained $\partial |B|/ \partial z =$ \SI{59.7(1)}{\tesla\per\metre}.

\subsection*{SERS-driven quadrupole}

\begin{figure}[h]
\centering
\includegraphics[width=0.9\columnwidth]{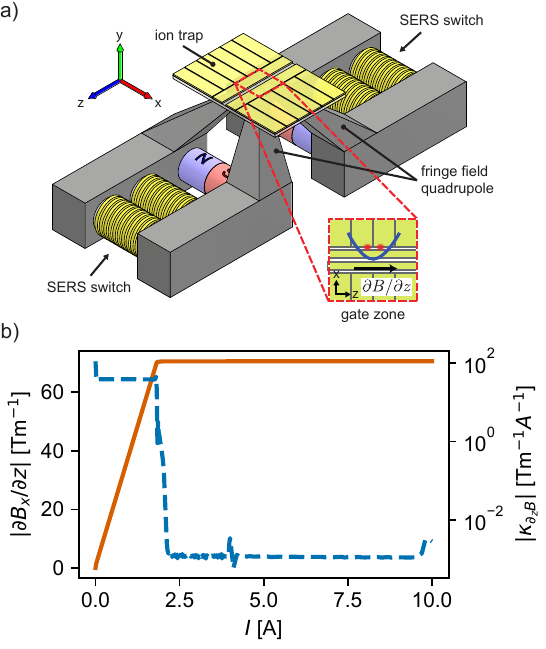}
\caption{a) Schematic of a SERS-driven fringe field quadrupole core prototype for creating switchable ultra-low noise magnetic field gradients in the gate zone of a trapped-ion quantum information processor. Here, the device is illustrated positioned underneath a gate zone region of surface trap, creating a field-gradient along the axial z-axis with quantisation direction along the x-axis. b) FEM simulation results of the magnetic field gradient magnitude produced at an ion height of \SI{500}{\micro\metre} above the core surface as a function of driving current (orange). Also plotted is the effective gradient sensitivity to current, $\kappa_{\partial_{z}B}$, as a function of applied current (blue, dashed).}
\label{fig-SERS-Q}
\end{figure}

Combining the SERS circuit with an RCA core, we present in Fig. \ref{fig-SERS-Q}(a) an example of a switchable low-noise magnetic-field gradient device. The core is designed to be located beneath the gate-zone of a surface ion trap, generating a large axial B-field gradient, $\partial_{z}B=\partial|B|/\partial z$, and field nil at the centre of the gate zone. We simulate the field at $h=$ \SI{500}{\micro\metre}, corresponding to a surface trap thickness of \SI{375}{\micro\metre} and ion trapping height of \SI{125}{\micro\metre} (see Methods). The core is magnetised by two SERS circuits.

In the OFF-state, $I=0$, a minimal remnant gradient of  $\partial_{z} B=$ \SI{310.4(9)}{\milli\tesla\per\metre} is obtained, enabling ion-transport across the gate zone. If required, each pair of shunts may be driven symmetrically,  $\eta_{1}=\eta_{2}$, such that they fully cancel the magnet's remnant field. Fig. \ref{fig-SERS-Q}(b) shows the obtained axial gradient as a function of applied current. With a \SI{2.3}{\ampere} switching current, an operational current of \SI{2.5}{\ampere} creates a \SI{200}{\milli\ampere} buffer to protect against ringing, offsets and drifts in the applied current. In the ON-state, a current-insensitive gradient of $\partial_{z}|B|=$ \SI{70.5(1)}{\tesla\per\metre} is obtained over a \SI{100}{\micro\metre} span, enabling two-qubit quantum gates. 

To determine the impact of the reduced gradient current sensitivity,
\begin{equation}
    \kappa_{\partial_{z}B}=\frac{d}{dI}\frac{\partial |B|}{\partial z},
\end{equation}
we make a direct comparison to CCW structures for switchable gradient generation. Siegele-Brown \textit{et al.} report CCWs with an effective $\kappa_{\partial_{z}B}=$ \SI{11.1}{\tesla\per\metre\per\ampere} for the same ion height of \SI{125}{\micro\metre} \cite{siegele2022}. Since CCW gradient strength scales linearly with current, a large $\kappa_{\partial_{z}B}$ is desired to achieve the required gradient strengths without prohibitively large Joule losses. However, this simultaneously increases the impact of current noise and drift. The SERS quadrupole displays a ramping $\kappa_{\partial_{z}B}$ of \SI{38.2(1)}{\tesla\per\metre\per\ampere}, followed by a sharp drop to a finite $\kappa_{\partial_{z}B}$ of \SI{1.2(2)E-3}{\tesla\per\metre\per\ampere} at saturation. These results indicate the device exhibits a larger efficiency than CCWs while attenuating current noise by up to five orders of magnitude. While $\kappa_{\partial_{z}B}$ will depend on fabrication tolerances, as shown earlier, orders-of-magnitude reductions in current sensitivity are still achievable in a non-ideal device. 

The ON-state continuous Joule losses are calculated as \SI{2.05}{\watt} at \SI{300}{\kelvin} (see Methods). For the same gradient, this is an order of magnitude reduction from the \SI{17.9}{\watt} estimated for CCWs at room temperature. The increased efficiency compared to CCWs is due to the ability to use multi-turn solenoids, as opposed to embedded CCWs being limited to a single layer. By shaping the magnetic field through its geometry, the FFQ core provides freedom to optimise the magnetising source for reduced power dissipation, inductance and manufacturability. Furthermore, the added distance between the dissipative solenoids and the trap surface provided by the device reduces heat transfer to the gate zone, reducing motional decoherence \cite{Bruzewicz2015}.   

\section*{Conclusion and outlook}\label{sec4}

The analysis and modelling presented in this work indicates that the SERS circuit can offer current-controlled magnetic-flux gating while strongly suppressing sensitivity to noise in the applied current. Experimental realisation of a SERS device will allow the remaining material-dependent limits, including magnetisation noise, drift and hysteresis-related repeatability to be quantified in practical devices. If coupled with a sufficiently low-noise current source, a SERS circuit could provide a switchable magnetic field with stability limited by the noise and drift properties of the employed magnetic materials, offering a useful tool for experimental research. Conversely, a near-ideal realisation of the circuit may substantially reduce the noise specification of power supplies used to generate magnetic fields, effectively acting as a passive broadband filter on the control current. Thus, having the potential to reduce the cost, complexity and environmental impact of such experiments. This work effectively shifts the problem of compromising between stability and bandwidth to one of manufacturing precision. Furthermore, SERS may provide a route to flux-gating fields from semi-hard magnets, superconducting magnets, or solenoid magnets in applications that are compatible with field delivery via a ferromagnetic core. For example, in persistent-current superconducting magnets, SERS could provide a route to modulating the delivered field without directly switching the superconducting current path to interrupt the persistent currents, such as with persistent-current-switch approaches \cite{Zhang2025}. 

The two presented FFQ gradient cores exploit the fringe effect to project a quadrupolar magnetic field into a volume of space outside of the core. By eliminating the need to interpose the experimental region in the core air-gap, they provide hindrance-free miniaturisation and are readily integrated into experiments. Hence, applicable to fields where large magnetic field gradients are required in a miniaturised device, such as magneto-optical traps \cite{Chen2022}, small-scale NMR imaging \cite{anders2018, Meissner2022,He2025}, electrical manipulation and individual addressing of spin qubits \cite{watson2018, Aldeghi2025}, and microfluidics \cite{Alnaimat2018, Descamps2021}. 

Combining the SERS circuit with an FFQ core, we provide a solution to one of the key challenges facing trapped-ion quantum computing with MAGIC on a QCCD architecture. The five orders of magnitude decrease in current sensitivity compared to CCWs creates a path towards permanent magnet level stability while remaining suitable for ion-transport. In addition, we calculate similar room-temperature power dissipation as that reported by CCWs at cryogenic temperatures \cite{siegele2022}. This relaxes the thermal management requirements for a large-scale computer, leading to reduced running costs and environmental impact.

\section*{Methods}\label{sec5}

\subsubsection*{Two-shunt SERS circuit equation}

A more rigorous determination of the air-gap induction as a function of $I$ is obtained by modelling the magnetic circuit through Hopkinson's law, 
\begin{equation}
\mathcal{F}=\mathcal{R}\Phi,
\label{SERSeq5}
\end{equation}
where $\mathcal{F}$ is the drop in magnetomotive force (\textit{mmf}) over a reluctance carrying flux $\Phi$, and Gauss's law,
\begin{equation}
\sum_{i}\Phi_{i}=0,
\label{SERSeq6}
\end{equation}
where $\Phi_{i}$ is the total flux carried by each element, $i$, between the primary core poles. The circuit consists of five parallel elements: the shunts (labels `$1$' and `$2$'), the magnet (labelled `$m$'), the air-gap (labelled `$g$') and the parallel flux leakage between the poles of the primary core (labelled `$\textrm{leak}$'). Examining the circuit in Fig. \ref{fig:sers-diagram}(b), since we assume the reluctance of the primary core to be negligible, there will be no significant \textit{mmf} drop between the parallel elements. As such, the shunts, magnet and air-gap all share a common \textit{mmf}, $\mathcal{F}_{g}$. The magnet contributes an \textit{mmf}, $\mathcal{F}_{m}$, to the circuit with series reluctance $\mathcal{R}_{m}$. Each shunt solenoid will contribute an \textit{mmf} of $\eta_{n}NI$ to the circuit, where  $\eta_{1}=-1$ and $\eta_{2}=1$. The flux in each shunt core will depend non-linearly on the $H$-field in the shunt, such that $\Phi_{1,2}=B(H_{1,2})A_{s}$. We also assume a non-negligible $\mathcal{R}_{\textrm{leak}}$. Through equation (\ref{SERSeq5}) and defining ``up'' as the positive direction for \textit{mmf} and flux, we find 

\begin{equation}
\Phi_{g} = -\frac{\mathcal{F}_{g}}{\mathcal{R}_{g}}, 
\end{equation}

\begin{equation}
\Phi_{\textrm{leak}} = -\frac{\mathcal{F}_{\textrm{g}}}{\mathcal{R}_{\textrm{leak}}}
\end{equation}
and
\begin{equation}
\Phi_{m} = \frac{\mathcal{F}_{m}-\mathcal{F}_{g}}{\mathcal{R}_{m}}.
\end{equation}
for the air-gap, leakage and the magnet respectively. We note that $\mathcal{F}_{g}$ is negative since it will generally act in opposition to the magnet. Given that, from equation (\ref{SERSeq6}) 

\begin{equation}
\sum_{i}\Phi_{i}=\Phi_{1}+\Phi_{2}+\Phi_{\mathrm{m}}+\Phi_{\textrm{leak}}+\Phi_{g} =  0,
\label{meth_gauss}
\end{equation}
 a governing two-shunt SERS circuit equation can be given as 
\begin{multline}
B(H_{1})A_s + B(H_{2})A_s \\
+\frac{\mathcal{F}_{m}-\mathcal{F}_{g}}{\mathcal{R}_{m}} - \frac{\mathcal{F}_{g}}{\mathcal{R}_{\textrm{leak}}}-\frac{\mathcal{F}_{g}}{\mathcal{R}_{g}} = 0,
\label{SERSeq7}    
\end{multline}
where from Ampere's law
\begin{equation}
H_{1}=\frac{-NI-\mathcal{F}_{g}}{\ell_{s}} \quad \textrm{and} \quad
H_{2}= \frac{NI-\mathcal{F}_{g}}{\ell_{s}} 
\label{SERSeq8}
\end{equation}
for a shunt length of $\ell_{s}$. Given the circuit parameters and the non-linear $B(H)$ function for the shunt material, $\mathcal{F}_{g}$ can be solved numerically for each value of applied current. The magnitude of the magnetic flux density in the air-gap is then given as $|B_{g}(I)| = \mu_{0}|\mathcal{F}_{g}(I)|/\ell_{g}$. 
 
\subsubsection*{SERS switching current}

To find the switching current, $I_{\text{sat}}$, we differentiate equation (\ref{SERSeq7}) with respect to $I$, yielding
\begin{multline}
A_s\frac{dB(H_{1})}{dH_{1}}\frac{dH_{1}}{dI} + A_s\frac{dB(H_{2})}{dH_{2}}\frac{dH_{2}}{dI} \\
- \left(\frac{1}{\mathcal{R}_{m}} +\frac{1}{\mathcal{R}_{\textrm{leak}}}+\frac{1}{\mathcal{R}_{g}}\right)\frac{d\mathcal{F}_{g}}{dI} = 0.
\label{SERSeq9}
\end{multline}

From equation (\ref{SERSeq8}) we obtain
\begin{multline}
\frac{A_s N}{\ell_{s}}\!\left(\frac{dB(H_{2})}{dH_{2}}-\frac{dB(H_{1})}{dH_{1}}\right) \\
-\frac{A_s}{\ell_{s}}\!\left(\frac{dB(H_{2})}{dH_{2}}+\frac{dB(H_{1})}{dH_{1}}\right)\frac{d\mathcal{F}_{g}}{dI}\\
-\left(\frac{1}{\mathcal{R}_{m}}+\frac{1}{\mathcal{R}_{\textrm{leak}}}+\frac{1}{\mathcal{R}_{g}}\right)\frac{d\mathcal{F}_{g}}{dI}
=0 .
\label{SERSeq10} 
\end{multline}
The on-state condition occurs when $d\mathcal{F}_{g}/dI = 0$, which is true when 
\begin{equation}
\frac{dB(H_{2})}{dH_{2}}= \frac{dB(H_{1})}{dH_{1}}.
\label{SERSeq11}
\end{equation}
This indicates that switching occurs when the differential permeabilities, $\mu'_{1,2}$, of each shunt are equal, i.e. becoming fully saturated such that $\mu'_{1}=\mu'_{2}=\mu_{0}$. We define $H_{\textrm{sat}}$ as the point on the shunt material's BH-curve where full saturation occurs. From equation (\ref{SERSeq8}) it is apparent that S2 saturates at higher current than S1. Therefore, switching occurs when $|H_{2}|=H_{\textrm{sat}}$, yielding
\begin{equation}
I_{\textrm{sat}} = \frac{1}{N}\left(H_{\textrm{sat}}\ell_{s}+\mathcal{F}_{g,\textrm{sat}}\right),
\label{SERSeq12}
\end{equation}
where $\mathcal{F}_{g,\textrm{sat}}$ is the air-gap \textit{mmf} at saturation, given by
\begin{equation}
\mathcal{F}_{g,\textrm{sat}} = \frac{\mathcal{R}_{\textrm{on}}\mathcal{R}_{\textrm{leak}}\mathcal{R}_{g}}{\Lambda}\mathcal{F}_{m},
\label{SERSeq13}
\end{equation}
where
\begin{multline}
\Lambda = \mathcal{R}_{\textrm{on}}\mathcal{R}_{\textrm{leak}}\mathcal{R}_{g}+\mathcal{R}_{m}\mathcal{R}_{\textrm{leak}}\mathcal{R}_{g} \\
+\mathcal{R}_{m}\mathcal{R}_{\textrm{on}}\mathcal{R}_{g}+\mathcal{R}_{m}\mathcal{R}_{\textrm{on}}\mathcal{R}_{\textrm{leak}}.
\end{multline}
Equivalently, $\mathcal{F}_{g,\textrm{sat}}$ can be expressed in terms of the air-gap flux density in the ON-state, $B_{g,\textrm{sat}}$, as $\mathcal{F}_{g,\textrm{sat}} = B_{g,\textrm{sat}}\ell_{g}/\mu_{0}$. $H_{\textrm{sat}}$ can be estimated from the shunt material's BH-curve while $\mathcal{F}_{g,\textrm{sat}}$ can be derived from the device geometry and magnet \textit{mmf} or through the required $B_{g,\textrm{sat}}$.

\subsubsection*{Numerical calculations for the example SERS device}

The example permanent magnet based SERS device is modelled with the parameters shown in Table \ref{tab-SERS-example-params}. The shunts and magnet are modelled as cylindrical with radii $r_{s}$ and $r_{m}$ respectively. The shunt-solenoids are assumed to have an inner radius and length equal to that of the shunts. We omit flux leakage in this example. Three representative material models are utilised from Comsol Multiphysics$^{\circledR}$'s materials library: nickel-steels ``Mu-metal$^{\circledR}$'' and ``4750'', and cobalt-steel ``Vanadium Permendur''. The normal (extrinsic) BH-curve is extracted from Comsol up to a field strength of \SI{1E6}{\ampere\per\metre} by simulating a perfectly wound closed-toroid of each material. This data is then interpolated in Python using SciPy's CubicSpline() function to obtain a callable $B(H)$ function. The differential permeabilities, $\mu'=dB/dH$, of each material are obtained using the first derivative of the cubic spline.

\begin{table}[]
\caption{Example permanent magnet based SERS device parameter list.}\label{tab-SERS-example-params}
\begin{tabular}{clll}
\hline
\multicolumn{1}{l}{Element} & Variable & Value & Description       \\ \hline\rule{0pt}{4ex}   
Shunts & \begin{tabular}[c]{@{}l@{}}$r_{s}$\\ $\ell_{s}$\\ $N$\end{tabular} & \begin{tabular}[c]{@{}l@{}}\SI{1}{\milli\metre}\\ \SI{4}{\milli\metre}\\ 30\end{tabular} & \begin{tabular}[c]{@{}l@{}}cylinder radius\\ cylinder length\\ turns\end{tabular}                               \\ \hline\rule{0pt}{4ex}
Magnet & \begin{tabular}[c]{@{}l@{}}$r_{m}$\\ $\ell_{m}$\\ $B_{r}$\\ $\mu_{r}$\end{tabular} & \begin{tabular}[c]{@{}l@{}}\SI{1}{\milli\metre}\\ \SI{4}{\milli\metre}\\ \SI{1}{\tesla}\\ 1.05\end{tabular} & \begin{tabular}[c]{@{}l@{}}cylinder radius\\ cylinder length\\ remanent flux density\\ relative permeability\end{tabular} \\ \hline\rule{0pt}{4ex}    
Air-gap & \begin{tabular}[c]{@{}l@{}}$A_{g}$\\ $\ell_{g}$\end{tabular}                       & \begin{tabular}[c]{@{}l@{}}\SI{4}{\square\milli\metre}\\ \SI{0.2}{\milli\metre}\end{tabular}& \begin{tabular}[c]{@{}l@{}}gap area\\ gap length\end{tabular}                                      \\ \hline
\end{tabular}
\end{table}

For the permanent magnet, the \textit{mmf} and reluctance are given by $F_{m}=\ell_{m}B_{r}/(\mu_{\mathrm{r}}\mu_{0})$ and $\mathcal{R}_{m}=\ell_{m}/\mu_{r}\mu_{0}A_{m}$, where $B_{r}$ is the magnet's remanent flux density and $\mu_{r}$ its relative recoil permeability. The air-gap flux density is given by $B_{g}(I) = \mu_{0}|\mathcal{F}_{g}(I)|/\ell_{g}$. We use SciPy's brentq() function, which implements Brent's root-finding method, to solve for $\mathcal{F}_{g}$ at each value of applied current in equation (\ref{SERSeq7}), given the model parameters and material $B(H)$ function. Leakage is neglected by assuming $\mathcal{R}_{\textrm{leak}}=\infty$, such that $\Phi_{\textrm{leak}}=0$. To find $\kappa(I)=dB_{g}(I)/dI$, we interpolate the obtained $B_{g}(I)$ using CubicSpline() and obtain the first derivative. Approximate prediction of the switching currents is carried out by visually estimating the $H$-field at which the differential permeability of each tested material reaches $\mu_{0}$. We estimate $H_{\mathrm{sat,\, MuMetal}}=$ \SI{400}{\ampere\per\metre},  $H_{\textrm{sat,\, NS \,4750}}=$ \SI{4E4}{\ampere\per\metre} and $H_{\textrm{sat,\, V\, Permendur}}=$ \SI{3E5}{\ampere\per\metre}. The switching current is then calculated using equation (\ref{SERSeq12}), where by taking $\mathcal{R}_{\textrm{leak}}=\infty$ equation (\ref{SERSeq13}) simplifies to 

\begin{equation}
\mathcal{F}_{g,\textrm{sat}} = \frac{\mathcal{R}_{\textrm{on}}\mathcal{R}_{g}}{\mathcal{R}_{\textrm{on}}\mathcal{R}_{g}+\mathcal{R}_{m}\mathcal{R}_{g}+\mathcal{R}_{\textrm{on}}\mathcal{R}_{m}}\mathcal{F}_{m}.
\end{equation}

\subsubsection*{Uncertainty estimation in finite element analysis}

When calculating magnetic field gradients and current sensitivities from results obtained by finite element analysis, two sources of uncertainty are considered: uncertainty in the accuracy of the simulated fields and uncertainty in the fitting of the field data. The former uncertainty is estimated by running the finite element simulation using two different mesh densities and comparing the difference between the obtained results. The highest mesh density used is limited by the computer's available memory. The smaller mesh density is chosen as a significant reduction in mesh elements that does not compromise convergence of the solution. Since we assume a further mesh refinement would lead to a smaller difference in the solutions than the one obtained, the obtained difference is used as an error bound. In the latter case, the uncertainty is the standard error obtained from the linear fit. The final reported uncertainty is the highest of the two.

\subsubsection*{Finite element analysis of the example SERS device}

\begin{figure}[h]
\centering
\includegraphics[width=0.9\columnwidth]{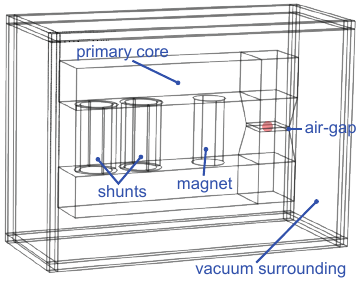}
\caption{Illustration of geometry used for the Comsol evaluation of an example SERS device. The air-gap flux density, $B_{g}$, is evaluated at the point marked by the red dot.}
\label{fig:sers-geometry-comsol}
\end{figure}

We carry out 3D non-linear finite element analysis of the example SERS device using Comsol's \textit{Magnetic Fields (mf)} interface of the AC/DC module. The model geometry is illustrated in Fig. \ref{fig:sers-geometry-comsol}. The device parameters are shown in Table \ref{tab-SERS-example-params}. Additionally, the primary core is formed of two rectangular cores of cross-section \SI{2.5}{\milli\metre} $\times$ \SI{2.5}{\milli\metre} and length \SI{12}{\milli\metre}. The primary core is made to taper into a \SI{2}{\milli\metre} $\times$ \SI{2}{\milli\metre} square air-gap. The primary core material is modelled using Comsol's non-linear Vacoflux 50, with a $B_{\textrm{sat}}$ of approximately \SI{2.3}{\tesla}. The shunts are modelled using Comsol's non-linear MuMetal. The shunt solenoids form a perfect cylindrical conductor around the shunt of thickness \SI{0.3}{\milli\metre}. The device is encapsulated in a box modelled with Comsol's ``perfect vacuum'' and infinite element domains are applied at the boundaries.   

Applied current is swept from \SI{10}{\ampere} to \SI{1}{\milli\ampere} using Comsol's \textit{Auxiliary Sweep} feature, facilitating convergence by starting in the fully-saturated regime. The air-gap B-field as a function of applied current is then evaluated at the geometric centre of the air-gap. Comsol's \textit{Magnetic Fields, No Currents (mfnc)} interface is used to evaluate the air-gap field at \SI{0}{\ampere}. By including leakage reluctance, $\mathcal{R}_{\textrm{leak}}$, in the numerical calculations, SciPy's curve\_fit() function is used to find the value of $\mathcal{R}_{\textrm{leak}}$ that minimises the residual between the analytical model and Comsol $B_{g}(I)$ data, yielding $\mathcal{R}_{\textrm{leak}}=$\SI{15.9}{\mega\per\henry}. 

\subsubsection*{SERS stability analysis}

In the ON-state, both shunt cores are fully saturated and their differential permeabilities are constant and equal to $\mu_{0}$. Under the assumption that only small perturbations, $\delta I$, in current occur, the resulting perturbation in air-gap \textit{mmf}, $\delta\mathcal{F}_{g}$, is described by a linear derivation of the magnetic circuit equation as
\begin{multline}
\frac{-N\delta I-\delta\mathcal{F}_{g}}{\mathcal{R}_{1}} + \frac{N\delta I-\delta\mathcal{F}_{g}}{\mathcal{R}_{2}} \\
+\frac{\mathcal{F}_{m}-\delta\mathcal{F}_{g}}{\mathcal{R}_{m}} - \frac{\delta\mathcal{F}_{g}}{\mathcal{R}_{\textrm{leak}}}-\frac{\delta\mathcal{F}_{g}}{\mathcal{R}_{g}} = 0.
\label{methEq-linSERS}
\end{multline}

 In this case, $\mathcal{R}_{1}=\ell_{s,1}/\mu_{0}A_{\textrm{sol,1}}$ and $\mathcal{R}_{2}=\ell_{s,2}/\mu_{0}A_{\textrm{sol,2}}$. In this case, $\ell_{s,1}$ and $\ell_{s,2}$ are the shunt lengths which are not assumed to be equal. We do, however, assume the solenoid lengths are equal to the shunt lengths to avoid flux leakage. Differentiating equation (\ref{methEq-linSERS}) with respect to $\delta I$ and given $\delta B_{g}=\mu_{0} \delta\mathcal{F}_{g}/\ell_{g}$ we obtain

 \begin{equation}
    \kappa_{\textrm{on}}=\frac{\mu_{0}\mathcal{R}_{t}N}{\ell_{g}}\left(\frac{1}{\mathcal{R}_{2}}-\frac{1}{\mathcal{R}_{1}}\right),
    \label{methEq-finitekappa}
 \end{equation}
where
\begin{equation}
\mathcal{R}_{t}=\left(\frac{1}{\mathcal{R}_{1}}+\frac{1}{\mathcal{R}_{2}}+\frac{1}{\mathcal{R}_{m}}+\frac{1}{\mathcal{R_{\textrm{leak}}}}+\frac{1}{\mathcal{R}_{g}}\right)^{-1}.
\end{equation}
Since $\mathcal{R}_{t}$ is in fact the effective total reluctance between the primary core poles, it will be almost invariant to small geometrical changes. Hence,  we can estimate $\mathcal{R}_{t}$ by assuming the ideal case in which 
\begin{equation}
\frac{1}{\mathcal{R}_{1}}+\frac{1}{\mathcal{R}_{2}}= \frac{1}{\mathcal{R}_{\textrm{on}}}.
\end{equation}

Given all geometric parameters, this expression allows us to calculate the finite $\kappa_{\textrm{on}}$ obtained when manufacturing errors prevent perfect satisfaction of condition D3. For $I<I_{\textrm{sat}}$ the OFF-state $\kappa_{\textrm{off}}=\mu_{0}N/\ell_{g}$, hence we can express the switching ratio as 
\begin{equation}
    \alpha=\left|\frac{\kappa_{\textrm{on}}}{\kappa_{\textrm{off}}}\right|=\mu_{0}\mathcal{R}_{t}\left|\frac{A_{\textrm{sol},2}}{\ell_{s,2}}-\frac{A_{\textrm{sol},1}}{\ell_{s,1}}\right|.
    \label{methEq-alpha}
\end{equation}
To evaluate flux leakage effects, we carried out the same Comsol simulation as detailed in the `Finite element analysis of the example SERS' section. Using Comsol's \textit{Parametric Sweep} feature, we reduce the length of one of the solenoids by 0\%, 1\% and 10\%. $\kappa_{\textrm{on}}$ is then evaluated by obtaining the first derivative of a CubicSpline() interpolation of the obtained $B_{g}(I)$ data. $\kappa_{\textrm{off}}$ is calculated analytically. The mean $\alpha$  obtained over the range $I=$ \SI{4.5}{\ampere} to \SI{7.5}{\ampere} is reported. With 0\% reduction, a baseline $\alpha=1.4\times10^{-5}$ is obtained. At 1\% and 10\% length reductions, $\alpha=1.9\times10^{-5}$ and $\alpha=6.9\times10^{-5}$ are obtained respectively. Being above baseline, these quantities indicate leakage related limits on $\alpha$. 

A finite primary core reluctance will result in additional series reluctance, $\delta\mathcal{R}$, between the shunt furthest from the air-gap (S1) compared to the one that is closest (S2). In an ideal device, we may model this added reluctance by setting $\mathcal{R}_{1}=\mathcal{R}+{\delta}\mathcal{R}$ and $\mathcal{R}_{2}=\mathcal{R}$, where $\mathcal{R}=\ell_{s}/\mu_{0}A_{s}$. Hence, from equation (\ref{methEq-finitekappa}), we find
\begin{equation}
\alpha=\mathcal{R}_{t}\left(\frac{\delta\mathcal{R}}{\mathcal{R}^{2}+\mathcal{R}\delta\mathcal{R}}\right).
\end{equation}
Therefore, it is important to ensure the primary core permeability remains large, i.e. it does not saturate, such that $\mathcal{R}\gg\delta\mathcal{R}$. For the example SERS device, we calculate $\delta\mathcal{R}$ by assuming a mean path length within the primary core of \SI{5.5}{\milli\metre},  cross-sectional area of \SI{2.5}{\square\milli\metre} and relative permeability $\mu_{r}=10^{4}$.
 
\subsubsection*{FFQ field calculations on Comsol}

Both the RCA and LCA fringe-field cores are simulated in 3D on Comsol using the \textit{Magnetic Fields, No Currents (mfnc)} interface of the AC/DC module. Both cores are modelled using Comsol's non-linear Vacoflux 50. The core poles are magnetised using two cylindrical permanent magnets with remanent flux density $B_{r}=$ \SI{1}{\tesla}, relative recoil permeability $\mu_{r}=$ 1.05, length $l=$ \SI{3}{\milli\metre} and radius $r_{m}=$ \SI{1}{\milli\metre}. The device is encapsulated in a box modelled with Comsol's ``perfect vacuum'' and infinite element domains are applied at the boundaries. The RCA core field-shaping parameters for optimal gradient magnitude at $h=$ \SI{500}{\micro\metre} are chosen based on results from previous work \cite{Taylor-Burdett2022}. For the LCA core, Comsol's \textit{Parametric Sweep} feature is used to find the field shaping parameters that create a magnetic field nil at $h=$ \SI{500}{\micro\metre}, without gradient magnitude optimisation. For each core $|B(x,y,z)|$ is evaluated along the z-axis at $x=0$ and $y=$ \SI{500}{\micro\metre}. The magnitude of the magnetic field gradient is then estimated through two separate linear regressions (SciPy's linregress() function) of $|B(z)|$ over two ranges, $z=$\SI{-15}{\micro\metre} to \SI{-115}{\micro\metre} and $z=$ \SI{15}{\micro\metre} to \SI{115}{\micro\metre}. Convergence of the result is checked through mesh refinement, the obtained results are shown in Table \ref{tab-FFQ-Comsol}.

\begin{table}[]
\caption{FFQ gradient evaluation, including mesh refinement, obtained by linear regression of $|B(z)|$ over different ranges. Standard error obtained from linear regression.}\label{tab-FFQ-Comsol}
\begin{tabular}{cccc}
\hline
Core                 & \begin{tabular}[c]{@{}c@{}}No. of \\ mesh elements\end{tabular} & \begin{tabular}[c]{@{}c@{}} $\partial_{z}B$ (\unit{\tesla\per\metre}) \\ {[}$z$-range (\unit{\micro\metre}){]}\end{tabular}     & \begin{tabular}[c]{@{}c@{}}Standard \\ error\\ (\unit{\tesla\per\metre})\end{tabular} \\ \hline\rule{0pt}{4ex}
\multirow{2}{*}{RCA} & 113,742                                                         & \begin{tabular}[c]{@{}c@{}}104.7 {[}-15, -115{]}\\ 104.7 {[}15, 115{]}\end{tabular} & \begin{tabular}[c]{@{}c@{}}0.4\\ 0.4\end{tabular}         \\ 
                     & 540,744                                                         & \begin{tabular}[c]{@{}c@{}}104.8 {[}-15, -115{]}\\ 104.7 {[}15, 115{]}\end{tabular} & \begin{tabular}[c]{@{}c@{}}0.4\\ 0.4\end{tabular}         \\ \hline\rule{0pt}{4ex}
\multirow{2}{*}{LCA} & 84,174                                                          & \begin{tabular}[c]{@{}c@{}}59.6 {[}-15,-115{]}\\ 59.6 {[}15, 115{]}\end{tabular}    & \begin{tabular}[c]{@{}c@{}}0.1\\ 0.1\end{tabular}         \\  
                     & 843,235                                                         & \begin{tabular}[c]{@{}c@{}}59.7 {[}-15, -115{]}\\ 59.7 {[}15, 115{]}\end{tabular}   & \begin{tabular}[c]{@{}c@{}}0.1\\ 0.1\end{tabular}         \\ \hline
\end{tabular}
\end{table}

\subsubsection*{SERS-driven FFQ prototype evaluation}

\begin{figure}[h]
\centering
\includegraphics[width=0.9\columnwidth]{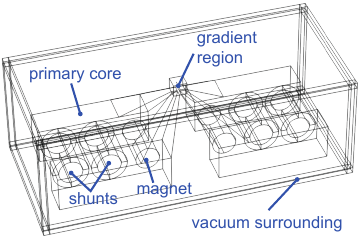}
\caption{Illustration of geometry used for the Comsol evaluation of the SERS-driven gradient device.}
\label{fig:sersQ-geometry-comsol}
\end{figure}

The permanent magnet SERS-driven gradient device is simulated on Comsol using 3D non-linear analysis with Comsol's \textit{Magnetic Fields (mf)} interface of the AC/DC module. The parameters of the shunts, magnets and solenoids are the same as detailed in Table \ref{tab-SERS-example-params}. The shunt solenoids form a perfect cylindrical conductor around the shunt of thickness \SI{0.5}{\milli\metre}. The primary core is formed of two rectangular cores of cross-section \SI{3}{\milli\metre} $\times$ \SI{2.5}{\milli\metre} and length \SI{9.3}{\milli\metre}. The RCA core is formed using the same pole geometry as detailed in Fig. \ref{core-geometries}. The primary core and RCA poles are modelled using the non-linear model for Vacoflux 50. The shunts are modelled using Comsol's non-linear MuMetal. Fig. \ref{fig:sersQ-geometry-comsol} illustrates the simulated geometry.

Obtained results are shown in Table \ref{tab-SERS-Q-Comsol}. Two meshing densities are simulated to ensure convergence of the simulated fields. For $x=0$ and $y=$\SI{500}{\micro\metre} the RCA core generates a field with $\partial B_{z}/\partial z= \partial B_{y}/\partial z = 0$ and $\partial_{z}B = |\partial |B|/\partial z| = |\partial B_{x}/\partial z|$. This facilitates calculation of $\partial_{z}B$ at each current by enabling a linear regression (using SciPy's linregress() function) from $z=$ \SI{-50}{\micro\metre} to \SI{50}{\micro\metre} of the $B_x(z)$ data at $x=0$ and $y=$ \SI{500}{\micro\metre}. $\kappa_{\partial_{z}B}$ is obtained from the first derivative of a cubic spline interpolation of the obtained $\partial_{z}B(I)$,  using SciPy's CubicSpline() function. The ramping $\kappa_{\partial_{z}B}$ is calculated from the mean of $\kappa_{\partial_{z}B}(I)$ in the range $I=$ \SI{0.5}{\ampere} to \SI{1.5}{\ampere}. For the ON-state $\kappa_{\partial_{z}B}$, the selected range is $I=$ \SI{2.5}{\ampere} to \SI{7.5}{\ampere}.

\begin{table}[]
\caption{Comsol simulation results for the SERS gradient device. For $\partial_{z}B$, standard error is obtained from the linear regression. For $\kappa_{\partial_{z}B}$, standard error is obtained from the set of data used to calculate the mean $\kappa_{\partial_{z}B}$. Two meshing densities are simulated: mesh 1 (41,003 elements) and mesh 2 (75,929 elements).}\label{tab-SERS-Q-Comsol}
\begin{tabular}{cccc}
\hline
Variable & \begin{tabular}[c]{@{}c@{}}Mesh\end{tabular} & Value    & \begin{tabular}[c]{@{}c@{}}Standard \\ error\end{tabular} \\ \hline\rule{0pt}{4ex}
\begin{tabular}[c]{@{}c@{}}$\partial_{z}B$  (\unit{\tesla\per\metre})\\ at $I=$ \SI{2.5}{\ampere}\end{tabular}  & \begin{tabular}[c]{@{}c@{}}1\\ 2\end{tabular}         & \begin{tabular}[c]{@{}c@{}}70.6\\ 70.5\end{tabular}     & \begin{tabular}[c]{@{}c@{}}0.1\\ 0.1\end{tabular}         
\\ \hline\rule{0pt}{4ex}

\begin{tabular}[c]{@{}c@{}}$\partial_{z}B$ (\unit{\milli\tesla\per\metre})\\ at $I=$ \SI{0}{\ampere}\end{tabular}  & \begin{tabular}[c]{@{}c@{}}1\\ 2\end{tabular}         & \begin{tabular}[c]{@{}c@{}}309.5\\ 310.4\end{tabular}   & \begin{tabular}[c]{@{}c@{}}0.4\\ 0.4\end{tabular}       \\ \hline\rule{0pt}{4ex}

\begin{tabular}[c]{@{}c@{}}$\kappa_{\partial_{z}B}$ ($\times10^{3}$)\\ at $I=$ \SI{2.5}{\ampere} \end{tabular} & \begin{tabular}[c]{@{}c@{}}1\\ 2\end{tabular}         & \begin{tabular}[c]{@{}c@{}}1.45\\ 1.24\end{tabular}       & \begin{tabular}[c]{@{}c@{}}0.03\\ 0.03\end{tabular}         \\ \hline\rule{0pt}{4ex}

\begin{tabular}[c]{@{}c@{}}$\kappa_{\partial_{z}B}$\\ at $I=$ \SI{1}{\ampere} \end{tabular} & \begin{tabular}[c]{@{}c@{}}1\\ 2\end{tabular}         & \begin{tabular}[c]{@{}c@{}}38.032\\ 38.161\end{tabular} & \begin{tabular}[c]{@{}c@{}}0.003\\ 0.003\end{tabular} \\ \hline

\end{tabular}
\end{table}

Joule losses are estimated assuming tight single-layer edgewise winding with rectangular wire. The wire is modelled as having a conductive cross-sectional thickness $a=$ \SI{0.1}{\milli\metre} and width $b=$ \SI{0.5}{\milli\metre} with a surrounding insulating layer of thickness $c=$ \SI{0.01}{\milli\metre}. The total conductive area is $A_{w}=ab$. Assuming a helical winding with average radius $r_{a}=r_{s}+b/2+c$ and pitch $p=\ell_{s}/N$, the total wire length is given by

\begin{equation}
L_w=N\sqrt{4\pi^{2}(r_{s}+b/2+c)^{2}+(\ell_{s}/N)^{2}}.
\end{equation}
The total Joule losses for all four solenoids are then given by
\begin{equation}
P = \frac{4\rho L_w{}}{A_{w}}I^{2},
\end{equation}
where $\rho=$ \SI{1.725E-8}{\ohm\metre} is the resistivity of copper at \SI{300}{\kelvin} \cite{Matula1979}. CCW Joule losses are estimated from their reported \SI{438}{\milli\ohm} resistance at room temperature and the current required to match the SERS quadrupole gradient given $\kappa_{\partial_{z}B}=$ \SI{11.1}{\tesla\per\metre\per\ampere} \cite{siegele2022}.

\section*{Acknowledgements}
\label{sec-Acknowledgements}
This work was supported by the U.K. Engineering and Physical Sciences Research Council via the EPSRC Hub in Quantum Computing and Simulation (EP/T001062/1), the U.K. Quantum Technology hub for Networked Quantum Information Technologies (No. EP/M013243/1), the European Commission’s Horizon-2020 Flagship on Quantum Technologies Project No. 820314 (MicroQC), the U.S. Army Research Office under Contract No. W911NF-14-2-0106 and Contract No. W911NF-21-1-0240, the Office of Naval Research under Agreement No. N62909-19-1-2116 and the University of Sussex. C.A.B. acknowledges support from the Engineering and Physical Sciences Research Council (EP/Y011864/1) via the Quantum Information Science \& Technologies Centre for Doctoral Training, Universal Quantum, and the University of Sussex.

\bibliography{references.bib}

\end{document}